\documentclass[10pt,conference]{IEEEtran}

\makeatletter
\let\MYcaption\@makecaption
\makeatother

\usepackage[font=footnotesize]{subcaption}

\makeatletter
\let\@makecaption\MYcaption
\makeatother

\usepackage{cite}
\usepackage{amsmath,amssymb,amsfonts}
\usepackage{algorithmic}
\usepackage{textcomp}
\usepackage{amssymb}
\usepackage{xcolor}
\def\BibTeX{{\rm B\kern-.05em{\sc i\kern-.025em b}\kern-.08em
    T\kern-.1667em\lower.7ex\hbox{E}\kern-.125emX}}
\usepackage{pifont}
\usepackage{graphicx}
\usepackage{url,hyperref,truncate}
\usepackage{float}
\hypersetup{colorlinks,allcolors=blue}
\graphicspath{ {./images/} }
\usepackage{balance}

\begin{document}

\title{OpenLambdaVerse: A Dataset and Analysis of Open-Source Serverless Applications}

\author{\IEEEauthorblockN{Ángel C. Chávez-Moreno}
\IEEEauthorblockA{\textit{Escuela Superior Politécnica del Litoral, ESPOL}\\
Guayaquil, Ecuador \\
acchavez@espol.edu.ec}
\and
\IEEEauthorblockN{Cristina L. Abad}
\IEEEauthorblockA{\textit{Escuela Superior Politécnica del Litoral, ESPOL}\\
Guayaquil, Ecuador \\
cabad@fiec.espol.edu.ec}
}

\maketitle

\begin{abstract}
    Function-as-a-Service (FaaS) is at the core of serverless computing, enabling developers to easily deploy applications without managing computing resources.
    With an Infrastructure-as-Code (IaC) approach, frameworks like the Serverless Framework use YAML configurations to define and deploy APIs, tasks, workflows, and event-driven applications on cloud providers, promoting zero-friction development.
    As with any rapidly evolving ecosystem, there is a need for updated insights into how these tools are used in real-world projects.
    Building on the methodology established by the Wonderless dataset for serverless computing (and applying multiple new filtering steps), OpenLambdaVerse addresses this gap by creating a dataset of current GitHub repositories that use the Serverless Framework in applications that contain one or more AWS Lambda functions.
    We then analyze and characterize this dataset to get an understanding of the state-of-the-art in serverless architectures based on this stack.
    Through this analysis we gain important insights on the size and complexity of current applications, which languages and runtimes they employ, how are the functions triggered, the maturity of the projects, and their security practices (or lack of).
    OpenLambdaVerse thus offers a valuable, up-to-date resource for both practitioners and researchers that seek to better understand evolving serverless workloads.
\end{abstract}


\maketitle

\section{Introduction}
\label{introduction}
Since the release of AWS Lambda in 2014, the Function-as-a-Service (FaaS) model has seen increasing adoption.
FaaS (serverless) solutions provide the tools to abstract infrastructure management away from the developer, while offering a pricing model based on usage, dynamically provisioning resources based on demand~\cite{Kounev:CACM}.
Current offerings enable the implementation of complex event-driven architectures, with features such as state management, workflow orchestration, and integration with cloud services like databases, storage systems, messaging queues, AI/ML pipelines, IoT services, and mobile backends. 
These offerings include commercial providers like AWS and Azure, and open-source alternatives like OpenWhisk and Kubeless; also available are tools that streamline development and deployment, like the popular Serverless Framework~\cite{serverlessframework}.
Despite its positives, serverless computing continues to present challenges that drive research, like dealing mitigating cold start latency, optimizing performance and cost-effectiveness for diverse workloads, addressing security concerns, and enabling interoperability.

To help focus research and development, others have collected and characterized applications that employ serverless functions~\cite{Eskandani:2021:Wonderless,Eismann:2022:Characterization,Bhatnagar:2023:OS3,Raffa:2023:Awsomepy}.
The Wonderless dataset \cite{Eskandani:2021:Wonderless, wonderlessdataset, wonderlessimplementation} consists of 1,877 projects from GitHub that use the Serverless Framework; the paper includes a brief characterization of the dataset, looking into programming languages, runtimes, and repository contributions.
The Open-Source Serverless Search (OS\textsuperscript{3})~\cite{Bhatnagar:2023:OS3} extended the application scraping with an improved approach and collection/curation code that excludes unlicensed and duplicate entries, and collected a dataset~\cite{os3dataset} with data from 5,981 repositories;
OS\textsuperscript{3} focuses on improved code scraping and supporting search, and does not include a characterization of the data.
AWSomePy~\cite{Raffa:2023:Awsomepy, awsomepydataset} also extended the Wonderless methodology to obtain a dataset with 145 serverless applications implemented in Python using the Serverless Framework for AWS Lambda; the characterization included analysis of plugins, code complexity, and cloud services and API usage.

With each new characterization effort, the community acquires a better understanding of how \emph{current} serverless applications are being built.
Our work updates and extends the Wonderless dataset, including updating the code so that it works with the new GitHub REST API and conforms to limit policies for code search and metadata collection\cite{githubapiratelimitdocs}.
\emph{OpenLambdaVerse} is the most up-to-date serverless application (code) dataset; the dataset is a collection of open-source serverless applications that use the Serverless Framework for AWS Lambda, written with a diverse set of programming language and runtimes.
We build upon the Wonderless methodology, and include the improvements introduced in OS\textsuperscript{3} and AWSomePy, providing key insights into the architecture and security policies on these applications, which could serve as reference for developers and researchers in the field.

The contributions from this work are:
\begin{enumerate}
    \item A new serverless repository extraction tool,\footnote{\url{https://github.com/disel-espol/openlambdaverse}} based on the Wonderless code plus improvements based on best practices of the repository mining community.
    \item An up-to-date, publicly available dataset\footnote{\url{https://zenodo.org/records/16533581}} of applications on GitHub that use the Serverless Framework for AWS.
    \item An analysis of the dataset showing current trends and a comparison with prior results, and insights from these. 
\end{enumerate}

\section{Dataset collection and curation} %
\begin{figure}[t]
        \centering
        \includegraphics[width=\columnwidth]{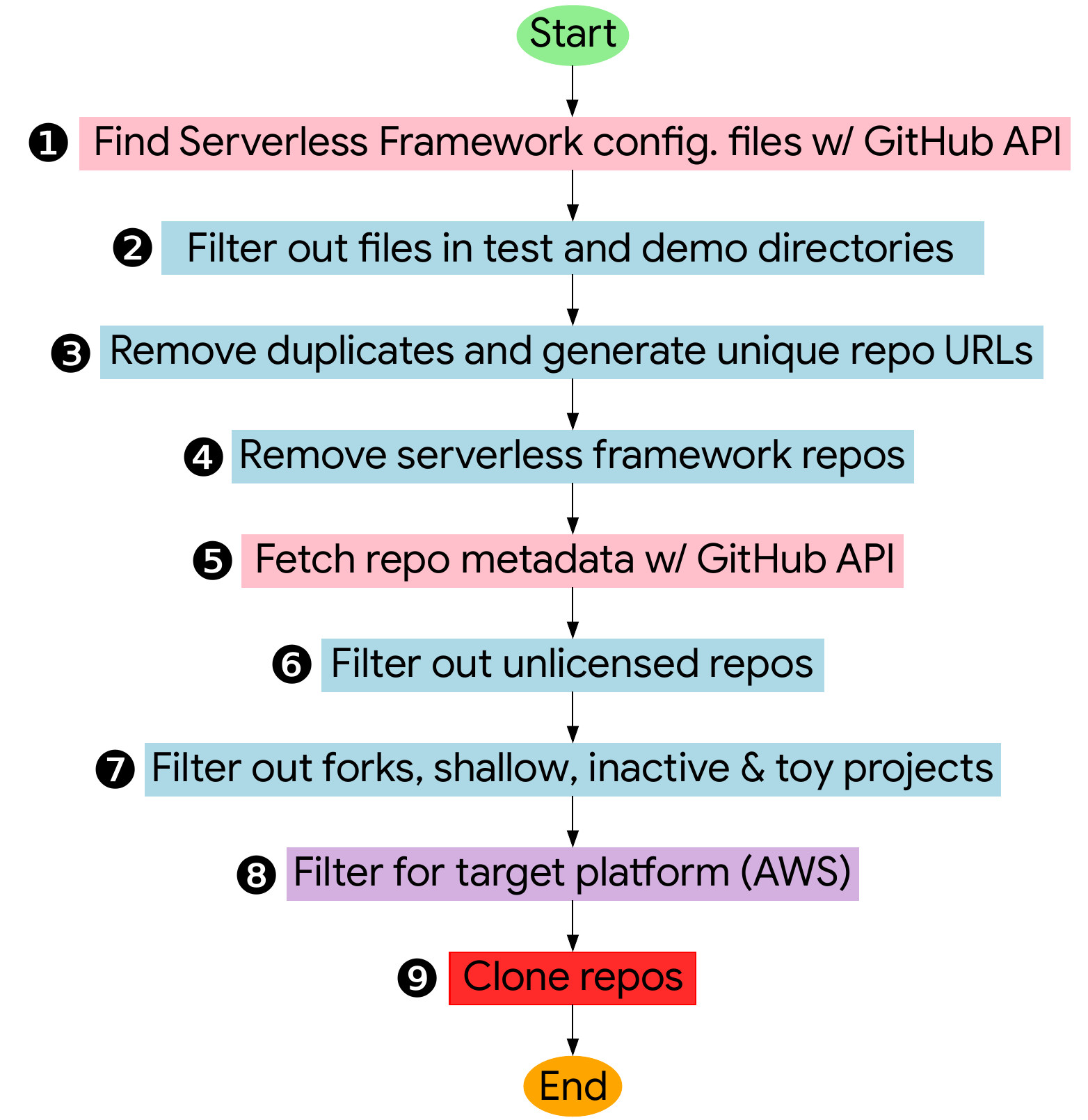}
        \vspace{-12pt}\caption{Repository extraction methodology. Steps in pink are where GitHub REST API is consumed. Filtering steps are in blue, except for the selection of AWS as target platform, in purple. Cloning of master branches, in red.}
        \label{fig:flowchart_en}
    \end{figure}

We ran our collection process on April 24, 2025, using the methodology outlined in Figure~\ref{fig:flowchart_en} and described below.

\newcounter{mycounter}
\setcounter{mycounter}{202} %

\vspace{6pt}\paragraph*{\hspace{-18pt}\ding{\value{mycounter}\stepcounter{mycounter}}\;Configuration files}
As in Wonderless and AWSomePy, OpenLambdaVerse collects serverless applications that implement the Serverless Framework using at least one \textit{serverless.yml} file for configuration and deployment.
In this first step, we obtained 34,320 configuration file URLs.

\vspace{6pt}\paragraph*{\hspace{-18pt}\ding{\value{mycounter}\stepcounter{mycounter}}\;Filtering out test and demo directories}
We filter out 8,925 URLs because they were part of example, test or demo directories.
This is a step done on previous studies as well.

\vspace{6pt}\paragraph*{\hspace{-18pt}\ding{\value{mycounter}\stepcounter{mycounter}}\;Removing duplicates and generating unique repository URLs}
20,144 URLs remained after removing duplicates stemming from more than one \textit{serverless.yml} in the same repo.

\vspace{6pt}\paragraph*{\hspace{-18pt}\ding{\value{mycounter}\stepcounter{mycounter}}\;Removing Serverless Framework contributions}
We identified and filtered out 64 repositories belonging to two official Serverless Framework accounts (users: \textit{serverless} and \textit{serverless-components}), leaving us with 20,080 URLs.

\vspace{6pt}\paragraph*{\hspace{-18pt}\ding{\value{mycounter}\stepcounter{mycounter}}\;Fetching repository metadata}
We fetch metadata from each of the repositories by parsing the serverless.yml file and by using GitHub's REST API to obtain: plugins, runtimes, events, provider, size, forks, stars, topics, primary programming language, flags: [archived, disabled, (is a) fork], bytes per programming language, contributors, private vulnerability reporting, tags, last commit date, stars, watchers, open issues, and license.
We compile this in a JSONL, where each JSON object contains key-value pairs with a repository's metadata.

\vspace{6pt}\paragraph*{\hspace{-18pt}\ding{\value{mycounter}\stepcounter{mycounter}}\;Filtering unlicensed projects} 
As in OS\textsuperscript{3}~\cite{Bhatnagar:2023:OS3}, we exclude projects that aren't licensed.
An unlicensed repository can't be reused without the owner's permission.
This makes our scraping follow a more direct path to applications that are purposely made open-source, leaving us with 4,080 repositories.

\vspace{6pt}\paragraph*{\hspace{-18pt}\ding{\value{mycounter}\stepcounter{mycounter}}\;Filtering forks, shallow, inactive and toy projects}
Before collecting (cloning) the repositories, we filter some out based on criteria collected by surveying papers in the Mining Software Repositories (MSR) research community:
\begin{itemize}
 \item Forks. As forks are derived from an existing project they contain duplicated code and distort the results~\cite{sens2024large}.
 \item Shallow projects. We define shallow projects as those $<100$KB, a heuristic to filter trivial/toy projects~\cite{sens2024large}.
 \item Inactive projects. We define inactive projects as those that have not been updated in the last 24 months; they are filtered out because they may not represent \emph{recent} trends in application development~\cite{he2025pinning,baltes2018no,coelho2018identifying,lu2025open,flint2022pitfalls}.\footnote{ This criteria replaces the project maturity threshold of Wonderless (last -- first commit $\ge1y$), which could stem from the purpose of these projects; e.g., created as part of a competition. 
 However, most projects created on those contexts are filtered out with our current criteria (e.g. size), and we decided not to exclude recent projects as they have the potential to provide additional insights in the rapidly evolving serverless ecosystem.}
 \item Toy projects. We filter out projects that contain any of the following keywords in the name, description or topics: (`example', `tutorial', `demo', `sample', `starter', `playground', `hello-world', `test', `template', `learn', `workshop', `exercise', `skeleton', `boilerplate', `mock', `poc', `guide'). 
 This list was compiled from the code of Wonderless and AWSomePy, plus 5 keywords we added after an exploratory analysis (last 5 in the list).
\end{itemize}

These filters leave us with 816 repositories in our list. 

\vspace{6pt}\paragraph*{\hspace{-18pt}\ding{\value{mycounter}\stepcounter{mycounter}}\;Selecting AWS as the target platform}
We keep only the projects where AWS is the target platform (in serverless.yaml), as it is the only platform currently supported by the Serverless Framework.\footnote{For v4, the framework deprecated the support of non-AWS providers~\cite{serverlessFramework:2023:state}.}
This brings the final count to 668 repositories.

\vspace{6pt}\paragraph*{\hspace{-18pt}\ding{\value{mycounter}\stepcounter{mycounter}}\;Cloning the repositories}
Finally, we clone the main branch of the repositories to preserve the snapshots associated with the dataset.
The aggregate size of the repositories is 51GB and has been released compressed (6GB) on \href{https://zenodo.org/records/16533581}{Zenodo}.

\subsection*{Notes on the consumption of the GitHub REST API}
    Our scraping code includes functionality to conform to the limits of the current GitHub REST API version 2022-11-28 for authenticated (non-enterprise) users~\cite{github:rate-limit,github:search}, and the rate limit changes to the search API introduced in 2023~\cite{github:2023:CodeSearchAPI}: 
    \begin{itemize}
      \item Only files $<$384KB can be found via code search.\footnote{Wonderless doesn't include serverless.yml files $<$0.5KB as they tend to be empty. We do not exclude small files unless other steps exclude the project.} This should not affect results as large YAML files are uncommon (e.g., we found only 3 in 86KB-168KB range). %
      \item GitHub's API rate limit is 5,000 requests per hour; we use a sleep timer to avoid exceeding this limit.
      \item The code search endpoint has a rate limit of 10 reqs/min; to conform, we use 7-sec sleep timers between requests.
    \end{itemize}

\section{Dataset characterization}
We present an analysis of OpenLambdaVerse with respect to the year of creation, use of plugins, code complexity, runtimes (and programming languages), topics, repository sizes, popularity, event triggers and security-related metadata.

\subsection{Year of creation}
\begin{figure}[t]
        \centering
        \includegraphics[width=\columnwidth]{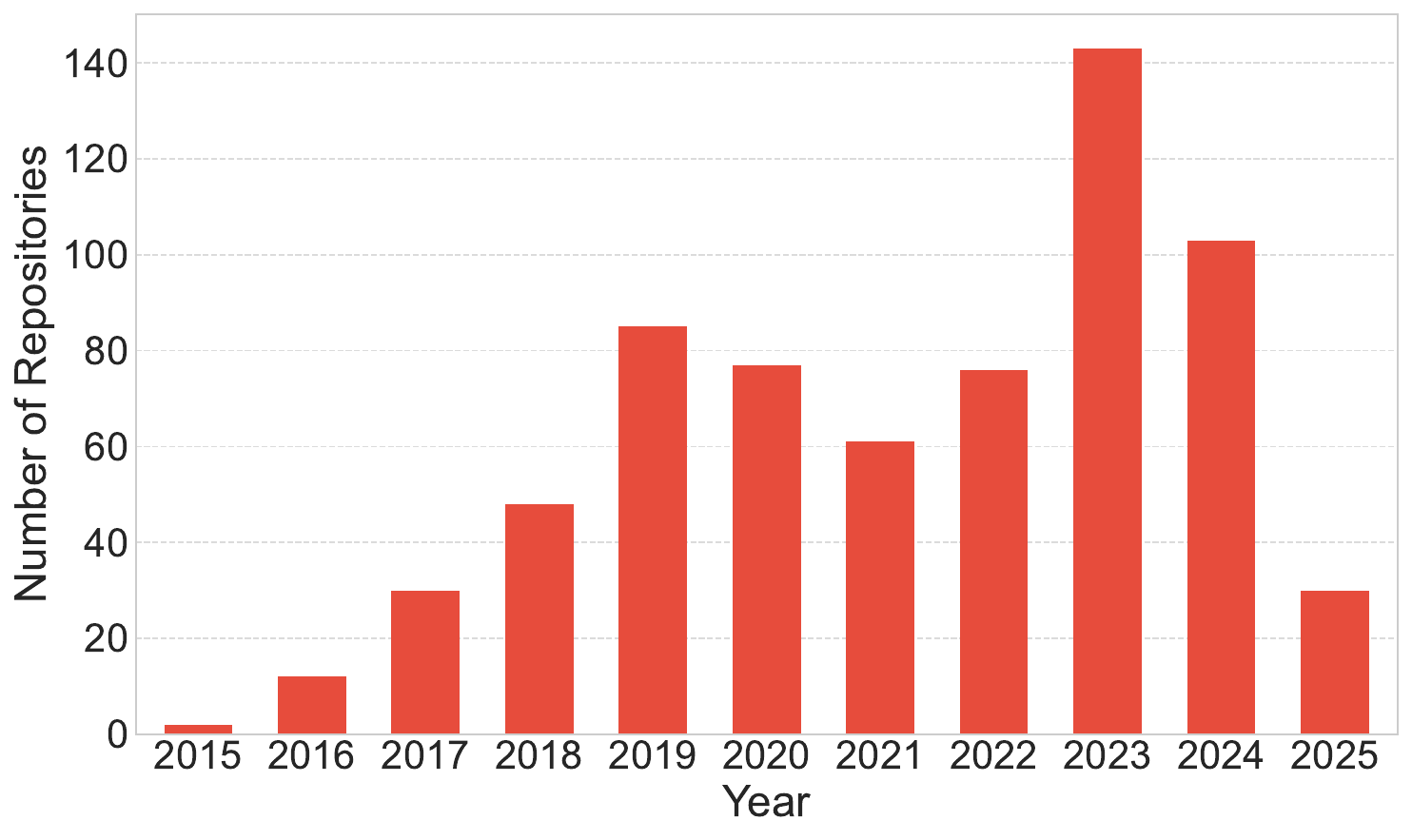}
        \vspace{-20pt}\caption{Creation year of the repos in OpenLambdaVerse (cutoff: 24/04/2025).}
        \label{fig:yearly}
    \end{figure}

As the Serverless Framework may be declining in popularity~\cite{datadog:2023:state}, we wanted to make sure that OpenLambdaVerse was not biased towards older repositories.
Fig.~\ref{fig:yearly} shows the number of repositories per year of creation.
A declining popularity of the framework is not yet observable, as the dip in repositories observed in 2025 corresponds to it being a partial year.

\label{eda}

\subsection{Plugins}
    \begin{table}[t]
    \caption{Top 11 Serverless Framework plugins.}
    \label{table:plugin_counts}
    \centering
    \begin{tabular}{|l|r|}
        \hline
        \textbf{Plugin} & \textbf{Count} \\ %
        \hline\hline
        serverless-offline & 270 \\
        \hline
        serverless-webpack & 91 \\
        \hline
        serverless-dotenv-plugin & 87 \\
        \hline
        serverless-plugin-typescript & 80 \\
        \hline
        serverless-python-requirements & 71 \\
        \hline
        serverless-prune-plugin & 49 \\
        \hline
        serverless-domain-manager & 42 \\
        \hline
        serverless-esbuild & 31 \\
        \hline
        serverless-bundle & 25 \\
        \hline
        serverless-iam-roles-per-function & 25 \\
        \hline
        serverless-dynamodb-local & 24 \\
        \hline
    \end{tabular}
    \end{table}
    
    The Serverless Framework provides plugins to extend its core functionalities.
    Table~\ref{table:plugin_counts} describes the top 11 plugins in the dataset and the number of repositories that use each.
    The leading plugin is \emph{serverless-offline}, which emulates AWS Lambda and API Gateway on a local machine; this is useful for prototyping while avoiding cloud usage costs.
    Next in popularity is \emph{serverless-webpack}, used to bundle Lambda functions using Webpack; this plugin simulates local API Gateway endpoints. Similarly for local testing is \emph{serverless-dynamodb-local}.
    \emph{serverless-dotenv-plugin} handles secrets in .env files, while \emph{serverless-plugin-typescript} and \emph{serverless-python-requirements} are needed when using Typescript and Python, respectively.
    \emph{serverless-iam-roles-per-function} is for access management in AWS.
    However, seeing that it is present in just 25 out of 668 projects (3.59\%), it seems that the best practices for applying governance are not used consistently (this was also observable in AWSomePy~\cite{Raffa:2023:Awsomepy}).

\subsection{Code complexity (measured through lines of code, LOC)}
    \begin{table}[t]
    \centering
    \caption{Lines of code per programming language.}
    \label{tab:language-stats-with-percentage}
    \begin{tabular}{|l|r|r|r|}
        \hline
        \textbf{Language} & \textbf{Files} & \textbf{LOC} & \textbf{Percentage (\%)} \\
        \hline\hline
        PHP & 25,454 & 6,163,832 & 46.15\% \\
        JavaScript & 16,522 & 5,081,974 & 38.05\% \\
        TypeScript & 22,880 & 1,713,025 & 12.83\% \\
        Python & 2,053 & 179,825 & 1.35\% \\
        Rust & 49 & 74,902 & 0.56\% \\
        Go & 698 & 57,893 & 0.43\% \\
        C\#\ & 269 & 19,805 & 0.15\% \\
        Cython & 43 & 16,243 & 0.12\% \\
        Java & 231 & 15,788 & 0.12\% \\
        Ruby & 27 & 915 & 0.01\% \\
        PowerShell & 2 & 160 & $<$0.01\% \\
        Other & 808 & 30,913 & 0.23\% \\
        \hline\hline
        \textbf{Total} & \textbf{69,036} & \textbf{13,355,275} & \textbf{100.00\%} \\
        \hline
    \end{tabular}
    \end{table}

    \begin{figure}[t]
    \centering
    \includegraphics[width=0.8\columnwidth]{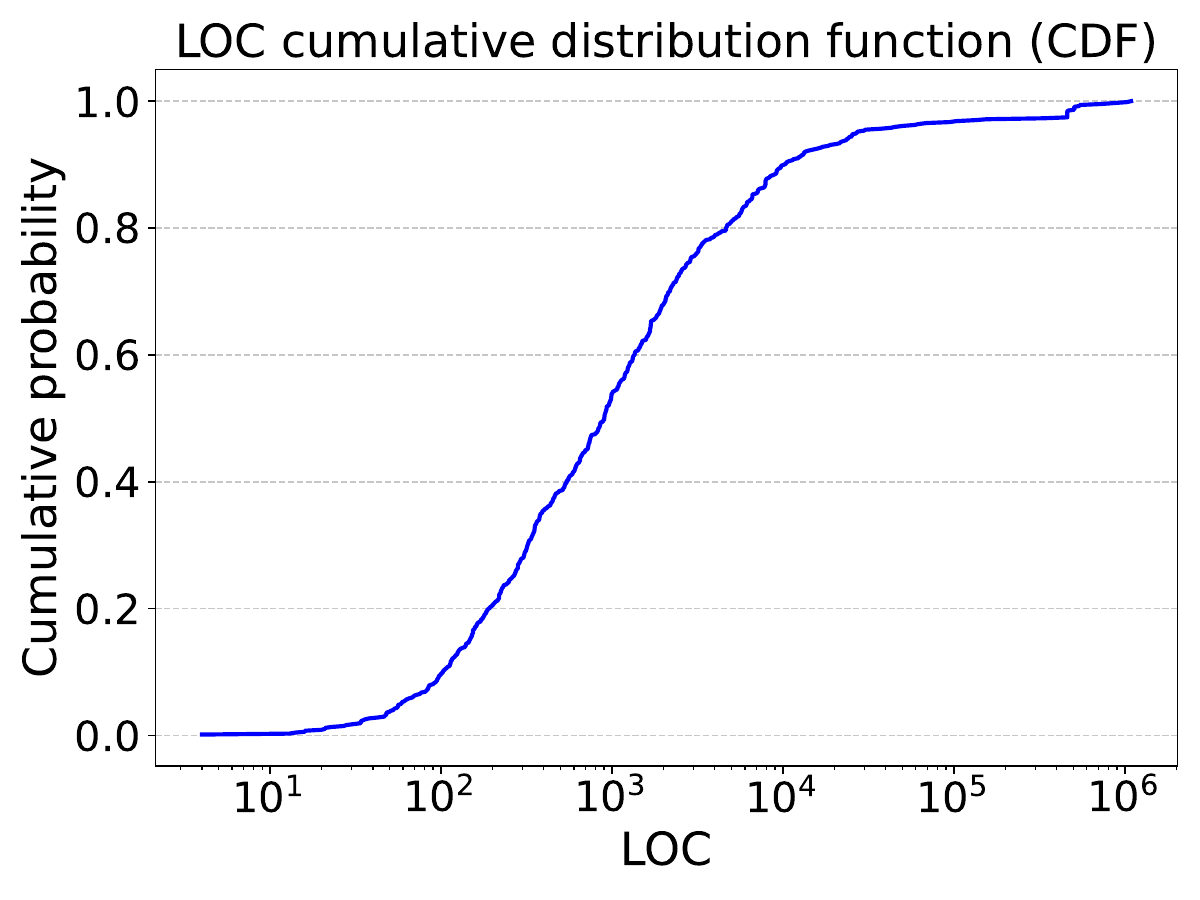}
    \vspace{-10pt}\caption{LOC CDF for all code in the repositories (not only function handlers).}%
    \label{fig:LOCs_log}
    \end{figure}

    We use CLOC~\cite{cloc} to analyze the lines of code (LOC) of the repositories and then analyze the results to find which are the most used programming languages in the dataset (shown in Table~\ref{tab:language-stats-with-percentage}).\footnote{To be consistent with the analysis of the bytes of code (III.D) done using GitHub's API, for III.C and III.D, we use the list of all languages known to GitHub, available at: \url{https://github.com/github-linguist/linguist/blob/main/lib/linguist/languages.yml} and select those with ``type: programming''. }
    Figure~\ref{fig:LOCs_log} shows the cumulative distribution function (CDF) of the lines of code in each repository, with a mean LOC of 20,022.
    Regarding the programming languages in the repositories (not just the function handlers), the leading one is PHP (46.1\%), followed by JavaScript (38.1\%), TypeScript (12.8\%) and Python (1.4\%), where the percentages are calculated as the LOC in that language to the total LOC in OpenLambdaVerse.
    Rust and Go are more popular than some traditionally popular languages like C\#, Java and Ruby, highlighting their recent rise in popularity and adoption.

\subsection{Bytes of code}

    \begin{table}[t]
    \centering
        \caption{Top programming languages by total code size (bytes).}
    \label{tab:language-bytes}
    \begin{tabular}{|l|r|r|}
        \hline
        \textbf{Language} & \textbf{Total Bytes} & \textbf{Occurrence (\%)} \\
        \hline\hline
        PHP & 362,887,947 & 52.36\% \\
        JavaScript & 247,084,664 & 35.65\% \\
        TypeScript & 64,327,098 & 9.28\% \\
        Python & 7,814,919 & 1.13\% \\
        Rust & 2,773,185 & 0.40\% \\
        Go & 1,962,996 & 0.28\% \\
        Cython & 904,887 & 0.13\% \\
        ANTLR & 857,375 & 0.12\% \\
        C\# & 830,584 & 0.12\% \\
        Java & 772,276 & 0.11\% \\
        C & 737,340 & 0.11\% \\
        Other & 2,163,226 & 0.31\% \\
        \hline\hline
        \textbf{Total} & \textbf{693,116,497} & \textbf{100.00\%} \\
        \hline
    \end{tabular}
    \end{table}

    The GitHub API has an endpoint to list the languages on a repository and the bytes of code written in that language~\cite{githubapiprogramminglanguages}.
    The aggregates per programming language are in Table~\ref{tab:language-bytes}.
PHP is not natively supported by AWS Lambda so the high prevalence (52.4\%) most likely comes from it being used to implement dynamic pages; however, some small percentage could come from handlers, as the OS-only (provided) runtime can be used to run unsupported languages in AWS Lambda.
    JavaScript (35.7\%) and TypeScript (9.3\%) are the most popular AWS Lambda-supported languages in the dataset.

    The high-presence of non-Lambda languages (bytes and LOC) makes it evident that there are limitations with an automated analysis of the programming languages present in the repositories, so to get a better picture of the prevalence of languages in serverless functions, it is best to analyze the runtimes used (next Subsection).

\subsection{Runtimes}
    \begin{table}[t]
    \centering
    \caption{Top runtimes used by the serverless functions.}
    \label{tab:merged-runtime-stats}
    \begin{tabular}{|l|r|r|}
        \hline
        \textbf{Runtime} & \textbf{Count} & \textbf{Percentage} \\
        \hline\hline
        nodejs & 426 & 73.58\% \\
        python & 94 & 16.23\% \\
        provided (OS-only) & 20 & 3.45\% \\
        java & 16 & 2.76\% \\
        go & 12 & 2.07\% \\
        ruby & 7 & 1.21\% \\
        dotnet & 4 & 0.69\% \\
        \hline
    \end{tabular}
    \end{table}

Table~\ref{tab:merged-runtime-stats} lists the most popular function runtimes. Nodejs (JavaScript and TypeScript) is the leading runtime, used in 73.58\% of the functions, followed by Python (16.23\%).
OS-only (provided) runtimes can be used for programming languages not directly supported by Lambda~\cite{aws:2025:runtimes}, like Go and Rust.
The unsupported Go appears in the table (2.07\%) because Lambda supported Go 1.x runtime up to Jan 2024~\cite{aws:2025:LambdaRuntimes}.

\subsection{Topics}

\begin{table}[t]
    \centering
    \caption{Top topics in the GitHub repositories.}
    \label{tab:topic-counts}

    \begin{tabular}{|l|r|}
        \hline
        \textbf{Topic} & \textbf{Count} \\
        \hline\hline
        \texttt{serverless} & 92 \\
        \texttt{aws-lambda} & 59 \\
        \texttt{aws} & 42 \\
        \texttt{serverless-framework} & 36 \\
        \texttt{lambda} & 36 \\
        \texttt{nodejs} & 32 \\
        \texttt{typescript} & 30 \\
        \texttt{cvs-project} & 17 \\
        \texttt{hacktoberfest} & 16 \\
        \texttt{dynamodb} & 15 \\
        \hline
    \end{tabular}
    \end{table}

    Topics are user-defined tags applied to GitHub repositories.
    Table~\ref{tab:topic-counts} shows the top 11 topics, including unsurprising ones like \emph{serverless}, \emph{aws-lambda}, \emph{aws}, \emph{serverless-framework} and \emph{lambda}.
    Popular topics also include runtimes or programming languages (\emph{nodejs} and \emph{typescript}) and a specific AWS service (\emph{dynamodb}).
    \textbf{\emph{hacktoberfest}} is a yearly competition hosted by GitHub; the projects left after filtering are those that are still active, highlighting the impact of this event.
    \emph{cvs-project} is the Commercial Vehicle Service (CVS) project from the UK's Driver \& Vehicle Standards Agency.

\subsection{Repository sizes}

    \begin{figure}[t]
        \centering
        \includegraphics[width=0.8\columnwidth]{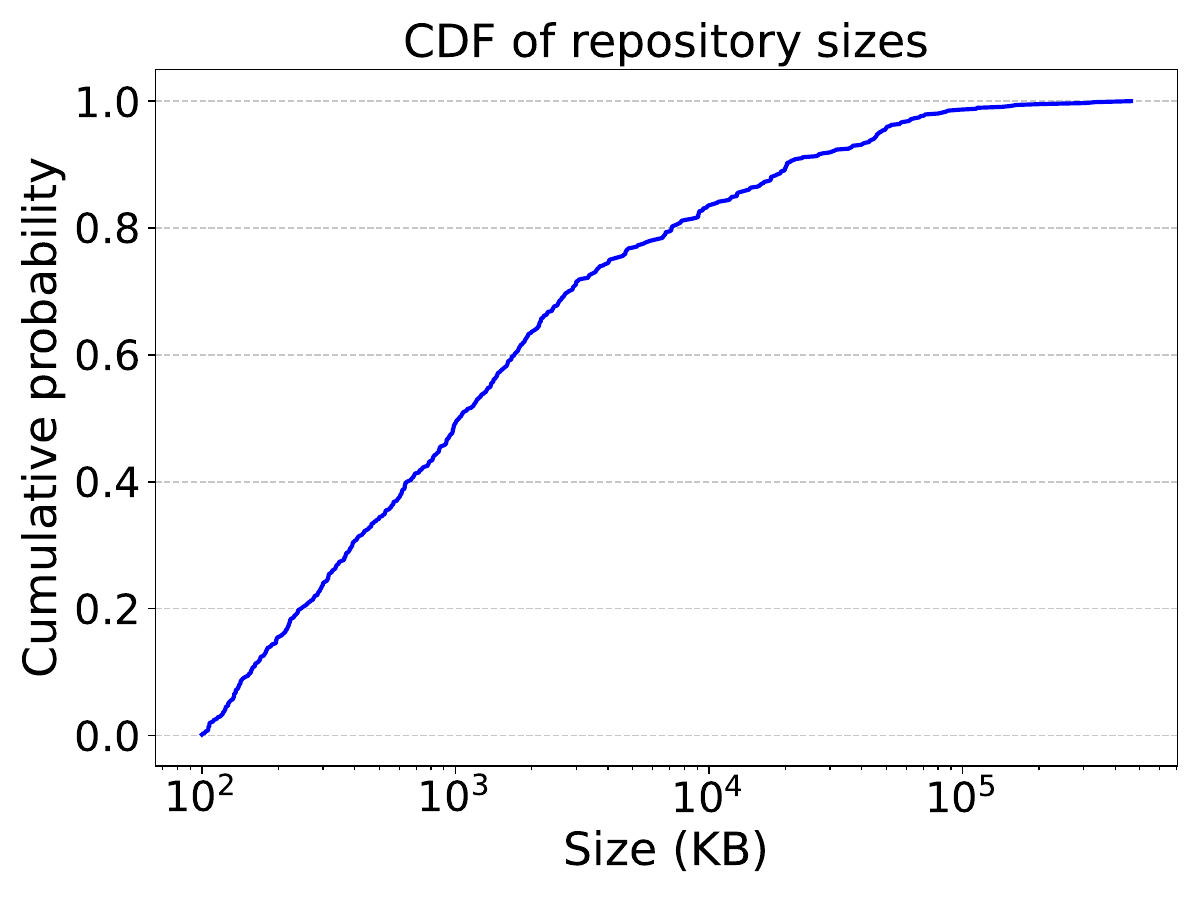}
        \vspace{-10pt}\caption{CDF of repository sizes.}
        \label{fig:cdf_repo_size_data_log}
    \end{figure}
    
    Figure~\ref{fig:cdf_repo_size_data_log} shows the CDF of repository sizes, which range from 100KB to 462MB, with a median of 1.03MB and a standard deviation of 31MB.

\subsection{Repository popularity}
    The number of contributors, forks, and issues serve as indicators of a project's popularity and adoption. OpenLambdaVerse repositories average 5 contributors per project, with 34.58\% of these repositories driven by a single contributor. The dataset also reveals an average of 7 open issues per project, which translates to notable community support. Although 43.71\% of the projects report zero open issues, the collected metadata alone does not suffice to conclude whether the projects initially had any open issues. Furthermore, a significant proportion (38.62\%) of these repositories have been forked at least once, another indicator of overall adoption.

\subsection{Event triggers}

    \begin{table}[t]
    \centering
    \caption{Event triggers used to call the serverless functions.}
    \label{tab:event-stats}
    \begin{tabular}{|l|r|r|r|}
        \hline
        \textbf{Event Trigger} & \textbf{Count} & \textbf{Triggers (\%)} &  \textbf{Repos (\%)} \\
        \hline\hline
        http & 1547 & 62.43\% & 51.05\% \\
        httpApi & 529 & 21.35\% & 23.20\% \\
        schedule & 199 & 8.03\% & 14.37\% \\
        websocket & 48 & 1.94\% & 1.80\% \\
        sqs & 30 & 1.21\% & 3.14\% \\
        sns & 27 & 1.09\% & 2.40\% \\
        s3 & 24 & 0.97\% & 2.69\% \\
        eventBridge & 17 & 0.69\% & 0.90\% \\
        stream & 11 & 0.44\% & 1.65\% \\
        cloudwatchEvent & 9 & 0.36\% & 1.05\% \\
        cognitoUserPool & 6 & 0.24\% & 0.45\% \\
        cloudwatchLog & 4 & 0.16\% & 0.45\% \\
        alexaSkill & 1 & 0.04\% & 0.15\% \\
        alb & 1 & 0.04\% & 0.15\% \\
        \hline
    \end{tabular}
    \end{table}

    Serverless code is launched in response to events like HTTP requests, scheduled tasks, and notifications.
    Table~\ref{tab:event-stats} lists these triggers and their prevalence in the dataset;
    \emph{http} and \emph{httpApi} are the most popular, representing 83.78\% of the triggers in the dataset, with  73.20\% of the repositories using one or the other at least once.\footnote{Both http and httpApi triggers are HTTP/REST triggers coming from API Gateway; the name is due to a version change: v2 (httpApi) or v1 (http)~\cite{serverless:2025:HttpApi}.}
    Scheduled events  constitute 8.03\% of triggers and appear in 14.37\% of the repositories.
    Triggers appearing in 2-4\% of the repositories are: \emph{sqs}, \emph{s3} and \emph{sns}.

    \begin{figure}[t]
        \centering
        \includegraphics[width=0.9\columnwidth]{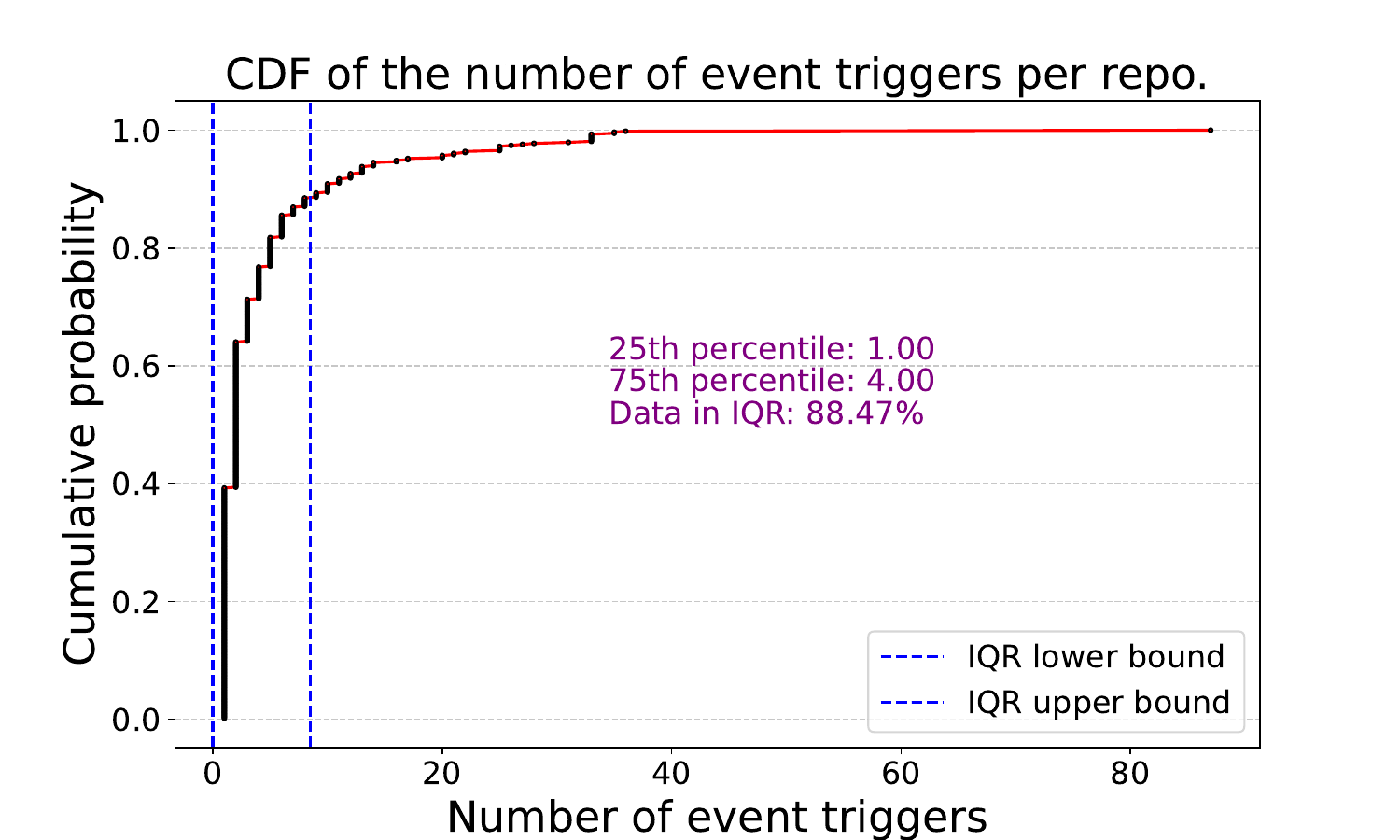}
        \vspace{-5pt}\caption{CDF of the number of event triggers per repository.}
        \label{fig:cdf_events_count_data}
    \end{figure}

    \begin{figure}[t]
        \centering
        \includegraphics[width=0.9\columnwidth]{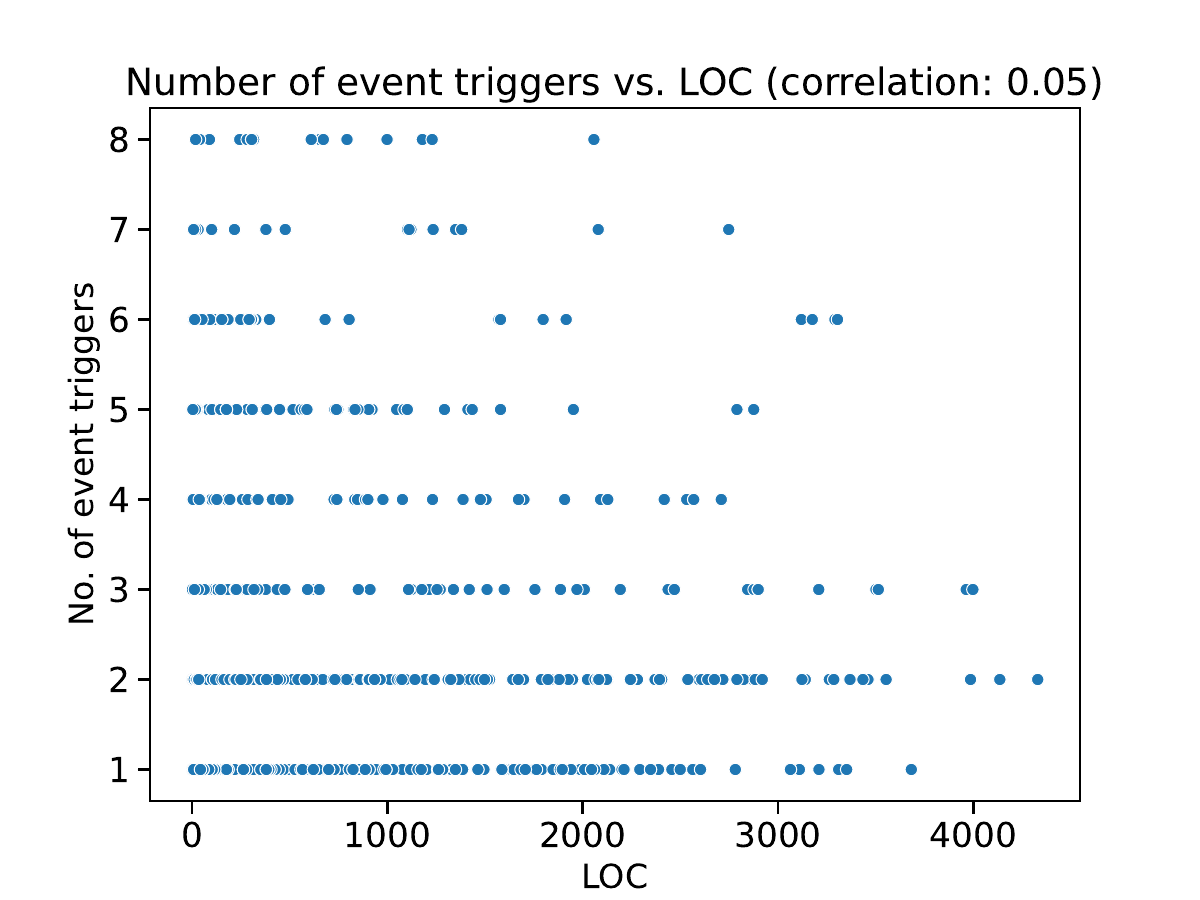}
        \vspace{-10pt}\caption{Number of event triggers vs. LOC}
    \label{fig:scatter_plot_event_triggers_vs_loc}
    \end{figure}
 
     Figure~\ref{fig:cdf_events_count_data} shows the CDF of the number of triggers per application; on the majority of the projects (88.47\%), the number of event triggers is between 1 and 4.
    As the number of triggers increases beyond that mark, we ask ourselves if this represents increased code complexity on function implementations.
    Figure~\ref{fig:scatter_plot_event_triggers_vs_loc} maps the relationship between the number of event triggers and code complexity (measured in LOC).
    The visual results and the 0.05 correlation coefficient suggest a negligible correlation between the variables.

\subsection{GitHub security-related metadata}
 Checking for private vulnerability reporting on a public repository is an additional key piece of metadata related to GitHub repository security~\cite{githubapiprivatevulnerabilityreportingendpoint}. This feature is crucial for constant vulnerability checks and updates. Seeing this implemented on only 98 repositories (14.67\%) reinforces our concerns regarding the lack of proper application of security best practices in serverless applications in the wild.

\begin{table*}[tb]
    \centering
    \scriptsize
    \caption{Serverless applications datasets and characterization.}%
    \label{tab:serverless:studies}
    \begin{tabular}{|l|c|p{8.3cm}|l|}
    \hline
    \textbf{Study \& Collection year} & \textbf{\#\,applications} & \textbf{Dataset description} & \textbf{Dataset URL} \\
    \hline\hline
    Wonderless~\cite{Eskandani:2021:Wonderless}\dotfill2020 & 1,877 & Repo URLs and cloned repositories (GitHub, Serverless Framework: AWS,
Azure, Google, OpenWhisk, Cloudflare, Kubeless). & \url{https://zenodo.org/records/4451387} \\ 
    Eismann et al.~\cite{Eismann:2022:Characterization}\dotfill2021 & 89 & Tabulated analysis of serverless applications sourced from GitHub, academic literature, and industrial literature; AWS, Azure, IBM, Google \& private. & \url{https://zenodo.org/records/5185055} \\
    AWSomePy~\cite{Raffa:2023:Awsomepy}\dotfill2023 & 145 & Dataset summary (metadata) and 145 cloned repositories (GitHub, Python, Serverless Framework: AWS). & \url{https://zenodo.org/records/7838077} \\
    OpenLambdaVerse\dotfill2025 & 668 & Tabulated metadata and cloned repositories (GitHub, Serverless Framework: AWS). & \url{https://zenodo.org/records/16533581} \\
    \hline
    \end{tabular}
\end{table*}

\section{Comparison with prior work}
\label{sec:comparison}
We discuss how our characterization compares to prior work in the serverless domain.
Table~\ref{tab:serverless:studies} lists the studies analyzed herein.
In addition to those studies, we also refer to the ``State of Serverless'' report by Datadog~\cite{datadog:2023:state}, which sources data from 20,000+ of companies in Datadog's customer base and was updated yearly from 2020 through 2023; it is not listed in the table because its source data is not publicly available.
We do not include trace-based studies that study/compile serverless calls without looking at the applications themselves (e.g., the ``Serverless in the Wild'' study~\cite{Shahrad:2020:ServerlessInTheWild}).

With the exception of the work by Eismann et al.~\cite{Eismann:2022:Characterization}, the other studies in Table~\ref{tab:serverless:studies} target the Serverless Framework due to its widespread adoption; this framework was an early product in this domain and it is widely used for deployment of serverless applications in AWS~\cite{Kritikos:2018:ReviewFrameworks,datadog:2023:state}.
Unlike the other studies in the table, Eismann et al. did a manual analysis of each of the 89 studied applications and this manual approach allowed them to consider any type of serverless application, at the cost of being the study that covered the smallest number of applications.
Wonderless also looked beyond AWS; however, for OpenLambdaVerse we did not consider other platforms as the Serverless Framework deprecated the support of non-AWS providers with the release of version 4 in May 2024~\cite{serverlessFramework:2023:state}.\footnote{\url{https://github.com/serverless/serverless/releases?q=4.0&expanded=true}}

Given that these studies target different frameworks and cloud providers, we cannot draw sound conclusions from comparing the results of the different characterizations of serverless applications.
Nevertheless, we do report how our results differ from prior studies, without making strong claims about what the differences in the results signify.

\paragraph*{Runtimes}
When compared to the stats in Wonderless, the popularity of the nodejs (73.6\% vs 72.2\%) and python (16.2 vs 19\%) runtimes has remained relatively stable, with nodejs increasing slightly in popularity and python decreasing in popularity.
These two runtimes are also reportedly the most common runtimes in the Eismann study (both at 42\%) and the Datadog study ($\sim$43\% nodejs and $\sim$28\% python), though the prevalence percentages differ due varied application inclusion methodologies.
In sum, all studies found that the most popular runtimes for serverless functions are nodejs and python, followed by java at a distant third place.

In OpenLambdaVerse and Wonderless, the other runtimes have popularities of less than 3.5\%, and have remained stable in the 5 years that have passed.
However, both Eismann and Datadog report Java runtimes having a higher prevalence (12\% and 10\%); this language also comes in third in popularity in OpenLambdaVerse and Wonderless, but at a lower popularity ($\sim$2.8\%). 
The difference in the Datadog numbers are particularly important given that they also report them for AWS only, so different provider support cannot explain them.
We believe this difference (higher popularity of Java in the Datadog study) may come from either inclusion of private repositories or the fact that they also included repositories with serverless applications managed by Terraform, which they report to be the preferred AWS Lambda deployment tool among larger organizations (which in turn may favor Java).

When comparing OpenLambdaVerse versus Wonderless we observe a $\sim$15\% decline in usage of Python and Go runtimes, the latter likely as a result of AWS deprecating this runtime in Jan 2024~\cite{aws:2025:LambdaRuntimes}.
The only runtime with a significant increase in adoption is ``provided'' ($\vartriangle$8\%) which may have seen increased adoption to support Go as this language is now unsupported by Lambda.
Datadog also found that custom runtimes are the fastest-growing type of functions and posit this may be due to an increased interest in serverless containers.

\paragraph*{Plugins}
AWSomePy is the only other study to include plug-in usage characterization.
We find the top three plug-ins to be 
serverless-offline (40\% prevalence),
serverless-webpack (14\% prevalence), and
serverless-dotenv-plugin (13\% prevalence).
These results differ from the AWSomePy results:
serverless-python-requirements (65\%),
serverless-pseudo-parameters (17\%), and
serverless-domain-manager (10\%),
presumably because AWSomePy only has Python applications.
In OpenLambdaVerse, serverless-python-requirements appears in 5\textsuperscript{th} place, with 10\% prevalence.

\paragraph*{Complexity (LOC)}
Wonderless and AWSomePy also report on the lines-of-code (LOC) per language and per repository, respectively.
However, our Table~\ref{tab:language-stats-with-percentage} is not comparable to Table 1 in the Wonderless paper~\cite{Eskandani:2021:Wonderless} because their analysis is done only for the handlers; thus, we reported a comparison between their Table 1 and our Table~\ref{tab:merged-runtime-stats} under the Runtimes heading of this Section.
With respect to the lines of code,  OpenLambdaVerse has average LOC per repository of 20,022, with 53.82\% of the repositories having 1K LOC or less, and 9.75\% having 100 LOC or less.
In AWSomePy the repositories are smaller on average (4,468 LOC), but have similar LOC for the 10 and 55 percentiles (the only percentiles specifically reported in~\cite{Raffa:2023:Awsomepy}): 55\% of the repositories have less than 1K LOC, and 10\% are under 100 LOC.

\paragraph*{Event triggers}
Only the Eismann (E) study looks at what types of triggers are used to trigger serverless functions.
In their study and in OpenLambdaVerse (O), the three most popular triggers are HTTP, Cloud events (sum of triggers coming from other service events like SQS, SNS and S3), and scheduled. 
However, the specific numbers differ:
HTTP (O: 86\%, E: 48\%),
scheduled (O: 16\%, E: 13\%),
cloud events (O: 12\%, E: 41\%).
These differences may not to represent changes in development practices but rather a result of different methodologies in selecting and analyzing the applications used by Eismann et al. (i.e., no automated scraping or analysis).

\section{Discussion}
\label{sec:discussion}

We found that serverless functions are frequently present in large projects that involve other, non-serverless components (e.g., PHP dynamic web pages), with functions written predominantly in node.js (JavaScript and Typescript) and Python, and triggered primarily by http and scheduled events.
We interpret this as an indicator that serverless functions are frequently used as part of larger, non-fully serverless projects, due to their simplicity and ease of implementation and deployment.

  \begin{figure}
    \includegraphics[width=\columnwidth]{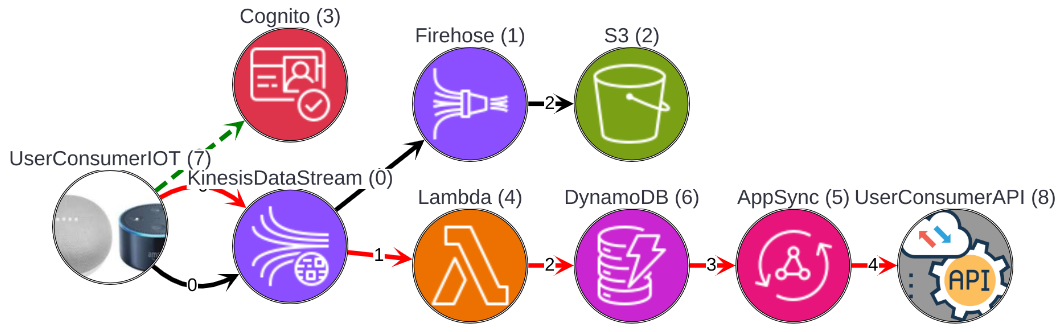}
    \caption{Sample IoT architecture from the Cloudscape dataset~\cite{Satija:2025:Cloudscape}: TechConnect (tech-powered sports with IoT machine learning).}
    \label{fig:iot}
  \end{figure}

Regarding the types of triggers used to call the Lambdas, we observe that IoT-specific triggers are missing, a surprising observation given that Lambda is used in a significant number of IoT/Edge applications:
A recent study from our group found that Lambda is present in 60\% of Edge architectures applications in a recent AWS-based dataset~\cite{Santillan:2025:HpcANDedge}.
However, the applications in that dataset may trigger Lambda through other mechanisms like http, stream (Kinesis), websockets, and indirectly through other cloud services. 
For an example of this, see Fig. 1 in~\cite{Santillan:2025:HpcANDedge} (reproduced here for clarity, in Fig.~\ref{fig:iot}).
In addition, it is possible that IoT applications on GitHub are, in a significant number of cases, closed-source or unlicensed and thus excluded by our filtering steps, or that these are preferring other IaC frameworks like Terraform~\cite{datadog:2023:state} or even other providers like Cloudflare.

The low rates of implementation of GitHub's private vulnerability reporting feature causes concern for the disclosure on critical vulnerabilities on these applications.
The owners of the repositories should implement tools that allow them to address security concerns privately before these become available to the public.
Further, not studied in this paper is the trustworthiness of plugins and libraries used in the projects, a growing concern in projects that depend on external packages~\cite{Alfadel:2023:npmVulnerabilities}; these security plugins should always be thoroughly evaluated for potential vulnerabilities before adoption.

\section{Threats to validity}
\label{threats_validity}
While we sought to be systematic in our data collection process and analysis, this study has limitations that could impact the generalizability and interpretation of our findings.
We categorize these threats into internal and external validity.

\subsection{External Validity}
\paragraph*{Scope and Representativeness of the Dataset}
Our analysis is based on publicly available repositories.
However, a substantial portion of software development, particularly for proprietary and enterprise-grade applications, occurs within private repositories.
The inaccessibility of this data limits the depth and breadth of our analysis, potentially introducing a bias towards practices prevalent in open-source projects. Therefore, the trends and characteristics identified in this study may not fully represent the complete landscape of serverless function usage, especially in production environments within organizations that heavily utilize private repositories.

\paragraph*{Technology Stack Specificity (Serverless Framework and AWS focus)}
A significant threat to the generalizability of our findings stems from our specific focus on repositories utilizing the Serverless Framework and targeting AWS Lambda.
The serverless ecosystem is diverse, containing diverse frameworks (including open source options like OpenWhisk) and cloud providers (e.g., Azure Functions, Google Cloud Functions, Cloudflare Workers), though AWS remains the provider with the largest serverless adoption~\cite{datadog:2023:state}.
Our reliance on the Serverless Framework for identifying serverless projects and AWS as the primary cloud provider means that our results might not be directly transferable to projects built with other frameworks or deployed on different cloud platforms;
for AWS, this framework remains the most popular infrastructure-as-code tool for managing AWS Lambda~\cite{datadog:2023:state} functions, but Terraform is more popular for highly-complex architectures~\cite{datadog:2023:state} so not considering Terraform biases our results.
Furthermore, new projects are increasingly preferring the AWS native options like CDK, SAM, and Chalice~\cite{datadog:2023:state}, and not including these in our study also limits the generalizability of our results.
Different frameworks offer varying features, conventions, and levels of abstraction, which could influence how serverless applications are structured, secured, and deployed.
Similarly, each cloud provider has unique services, best practices, and security models.
While AWS Lambda is a prominent serverless platform, a broader comparison including other frameworks or open-source platforms would enhance the generalizability of our findings.

\subsection{Internal Validity}

\paragraph*{Criteria and Limitations of Repository Filtering}
The accuracy of our findings depends on filters used to identify repositories that employ serverless functions (and that are not toy projects, demos, etc.).
To minimize this bias, we based our criteria in practices used by the mining software repositories community (and cited sources where appropriate), but this automated process might inadvertently exclude relevant repositories or include irrelevant ones.

\paragraph*{Interpretation of Security Practices from Code Analysis}
While our analysis highlighted concerns regarding the low usage of security plugins within the mined repositories, we acknowledge a problem with this approach:
Security features in cloud applications can be managed directly through the cloud provider's web interface.
Therefore, an absence of explicit security configurations within the code repository does not necessarily imply a lack of security best practices in the deployed application (though we can argue that not committing those configurations to the repositories is prone to manual error and is a security issue on itself).

\section{Conclusions}
\label{conclusions}
We presented OpenLambdaVerse, a dataset with 668 repositories obtained by using a scraping methodology based on best practices from the mining software repositories community and the original methodology proposed for the Wonderless dataset. 
Our characterization of the dataset goes significantly further than that of Wonderless and presents an up-to-date view of serverless applications in the wild.
We believe the dataset and characterization can be used by practitioners and academics to better understand how serverless workloads are evolving.
Furthermore, as we are releasing our dataset and code, others can build upon our work and continue to provide a more in depth understanding of the evolving serverless landscape.

\section*{Acknowledgments}
\label{sec:acknowledgements}
We want to thank our anonymous reviewers for their helpful suggestions that lead to an improved final version of this paper.

\balance
\bibliographystyle{IEEEtran}
\bibliography{bibliography.bib}

\end{document}